\documentclass[reprint,aps,prl,showpacs,amsmath,amssymb,floatfix,nolongbibliography]{revtex4-2}
\usepackage{amsthm}
\usepackage{amsmath, mathtools}
\usepackage[T1]{fontenc}
\usepackage[utf8]{inputenc}	
\usepackage[english]{babel}
\usepackage{amsmath}
\usepackage{graphicx}		
\usepackage{natbib}
\usepackage{textcomp}
\usepackage{gensymb}
\usepackage[usenames,dvipsnames,svgnames,table]{xcolor}
\usepackage[hidelinks,colorlinks=false,urlcolor=Cerulean,citecolor=black]{hyperref}
\usepackage{siunitx}
\usepackage{mathrsfs}
\usepackage{multirow}
\usepackage{bm}
\usepackage{xfrac}
\usepackage{comment}
\usepackage[normalem]{ulem}

\renewcommand{\i}{\mathrm{i}}

\newcommand{\revision}[1]{\textcolor{black}{#1}}

\begin{document}

\title{\revision{Theory of transient chimeras in finite Sakaguchi-Kuramoto networks}}

\author{Roberto C. Budzinski}
\author{James W. C. Graham}
\author{J\'an Min\'a{\v c}}
\author{Lyle E. Muller}
\affiliation{Department of Mathematics, Western University, London, ON, Canada}
\affiliation{Western Institute for Neuroscience, Western University, London, ON, Canada}
\affiliation{Western Academy for Advanced Research, Western University, London, ON, Canada}

\begin{abstract}
Chimera states are a phenomenon in which order and disorder can co-exist within a network that is fully homogeneous. \revision{Precisely how transient chimeras emerge in finite networks of Kuramoto oscillators with phase-lag remains unclear.} \revision{Utilizing an operator-based framework to study nonlinear oscillator networks at finite scale, we reveal the spatiotemporal impact of the adjacency matrix eigenvectors on the Sakaguchi-Kuramoto dynamics.} We identify a specific \revision{condition} for \revision{the emergence of transient} chimeras in these \revision{finite networks: the eigenvectors of the network adjacency matrix create a combination of a zero phase-offset mode and low spatial frequency waves traveling in opposite directions. This combination of eigenvectors leads directly to the coherent and incoherent clusters in the chimera.} \revision{This approach provides two specific analytical predictions:~(1) a precise formula predicting the combination of connectivity and phase-lag that creates transient chimeras, (2) a mathematical procedure for rewiring arbitrary networks to produce transient chimeras.}
\end{abstract} 

\maketitle

Chimera states are a central phenomenon in the study of nonlinear dynamics. Early computer simulations found that pockets of synchronous and asynchronous nodes could co-exist in a network where all nodes and connections were otherwise identical \cite{kuramoto2002coexistence}. This example prompted extensive theoretical work to understand how phase synchrony could spontaneously appear in a subset of nodes without systematic differences in either their intrinsic properties or their connectivity \cite{abrams2004chimera}. Since their initial discovery, chimera states have been observed in networks of phase oscillators \cite{montbrio2004synchronization,zhu2014chimera, abrams2008solvable,laing2012chimeras}, neural networks \cite{bansal2019cognitive,boaretto2021bistability,majhi2019chimera,santos2017chimera}, and other dynamical systems \cite{banerjee2016chimera,parastesh2021chimeras,dudkowski2014different,viktorov2014coherence}. Chimeras have also been found in experiments \cite{hagerstrom2012experimental,totz2018spiral,gambuzza2020experimental,patzauer2021self,hart2016experimental,tinsley2012chimera,wickramasinghe2013spatially}, demonstrating their existence in real-world systems and prompting further theoretical study of their underlying mechanisms.

The mathematical approaches that have been developed to study chimeras have been well suited to describe these states in continuous systems \cite{abrams2008solvable,omel2008chimera, kotwal2017connecting,nicolaou2017chimera}. The assumption of the continuum or asymptotic limit has made analytical insights possible in previous work, such as the identification of bifurcations \cite{abrams2004chimera,abrams2008solvable,laing2009chimera,kotwal2017connecting, panaggio2015chimera}. Further, numerical results have studied bifurcations leading to the emergence of chimera states \cite{bogomolov2017mechanisms,schulen2022solitary}. \revision{Several studies have analyzed the existence of weak chimeras, which are characterized by frequency synchronization \cite{ashwin2015weak,bick2016chaotic,bick2017robust}, and also explored these states in minimal networks \cite{maistrenko2017smallest,burylko2022symmetry}. Recent work has developed analytical approaches for strong chimera states in finite networks, where the synchronized and asynchronous clusters mutually interact to permanently stabilize the chimera in networks with two or more modules \cite{zhang2020critical,zhang2021mechanism}. Transient states, however, are increasingly understood to be important in neural dynamics and computation \cite{palmigiano2017flexible}, and chimeras are known to be transient states in finite networks of Sakaguchi-Kuramoto oscillators \cite{wolfrum2011chimera}. While techniques such as the master stability function or analysis of eigenvalues of the graph Laplacian have provided fundamental insight into the dynamics of oscillator networks, it is increasingly recognized that these techniques have limitations in studying these transient dynamics related to the brain \cite{papo2019brain}. At present, the underlying network mechanisms and specific mathematical conditions for the emergence of transient chimeras states in finite networks of Sakaguchi-Kuramoto oscillators remain unknown.}

Here, we report a \revision{analytical approach} for chimera states in finite networks of \revision{Sakaguchi-}Kuramoto oscillators. This \revision{mathematical approach} can be understood in geometric terms, where combinations of eigenvectors of \revision{the adjacency matrix} play \revision{a fundamental} role. Our results demonstrate that \revision{transient} chimera states can be understood as a combination of a zero phase-difference mode and low spatial frequency wave patterns traveling in opposite directions. This is valid for finite networks where different chimera patterns are observed. Our approach also identifies the specific roles played by connectivity and phase-lag between oscillators in generating \revision{transient} chimera states.

We use a recently introduced complex-valued\revision{, operator-based} approach to the Kuramoto model \cite{muller2021algebraic}, which is able to reproduce a diversity of synchronization patterns observed in Kuramoto networks \cite{budzinski2022geometry}. \revision{This approach} allows a direct connection between the eigenvectors describing the system and the shape of the resulting nonlinear dynamics \cite{budzinski2022geometry,budzinski2023analytical}. We start with the standard equation for the \revision{Sakaguchi-}Kuramoto network dynamics
\begin{equation}
    \dot{\theta}_i(t) = \omega + \epsilon \sum_{j=1}^{N} A_{ij} \sin\big( \theta_j(t) - \theta_i(t) - \phi\big)\,,
    \label{eq:km_phase_lag}
\end{equation}
where $\theta_{i}(t)$ is the phase of the $i^{\mathrm{th}}$ oscillator at time $t$, $\omega$ is the natural frequency, $\epsilon$ scales the coupling strength, $\bm{A}$ is the adjacency matrix, and $\phi$ is the phase-lag. \revision{Here, we consider a distance-dependent power-law connectivity, where the weights $A_{ij}$ decay with the distance between nodes $i$ and $j$ following a power-rule with periodic boundary conditions \cite{note_matrix}.} Without loss of generality, we consider oscillators with natural frequency \revision{of $\omega = 10\pi$}. We then introduce a complex-valued system that is related to the original, nonlinear Kuramoto model and allows connecting the eigenspectrum of the adjacency matrix to the resulting nonlinear dynamics \cite{muller2021algebraic,budzinski2022geometry,budzinski2023analytical}. Specifically, starting with unit-modulus initial conditions $|x_{i}(0)| = 1 \forall i$, with arguments $\mathrm{Arg}[x_{i}(0)]$ equal to the initial conditions of the Kuramoto system $\theta_{i}(0)$, we can use two operators to evolve a complex-valued system:
\begin{equation}
\bm{x}(t+\varsigma) = \Lambda \big[e^{\i \omega \varsigma} e^{\varsigma \bm{K}} \bm{x}(t)\big]\,,
\label{eq:lambda_operator_x}
\end{equation}
where $\varsigma$ is small but finite, $\Lambda$ represents an elementwise operator mapping the modulus of each state vector element $x_i$ to unity, and $\bm{K} = \epsilon e^{-\i \phi} \bm{A}$. We have shown in recent work that Eq.\,(\ref{eq:lambda_operator_x}) precisely reproduces the trajectories of the original Kuramoto system, specifically by comparing the variables $\theta_{i}(t)$ to $\mathrm{Arg}[x_{i}(t)]$ \cite{budzinski2022geometry}. Because all evolution of the arguments $\mathrm{Arg}[x_{i}(t)]$ is governed by the linear matrix exponential operator (with the elementwise $\Lambda$ operator only changing the moduli), we can analyze the nonlinear dynamics in this system in terms of the \textit{arguments} of the eigenvectors of $\bm{K}$ \cite{budzinski2023analytical,nguyen2023equilibria}. Thus, while connecting the eigenspectrum of a network's adjacency matrix to the resulting nonlinear dynamics is a difficult problem in general, here we demonstrate that this is possible in the \revision{Sakaguchi-}Kuramoto oscillator system.

\begin{figure}[b!]
    \centering
    \includegraphics[width=\columnwidth]{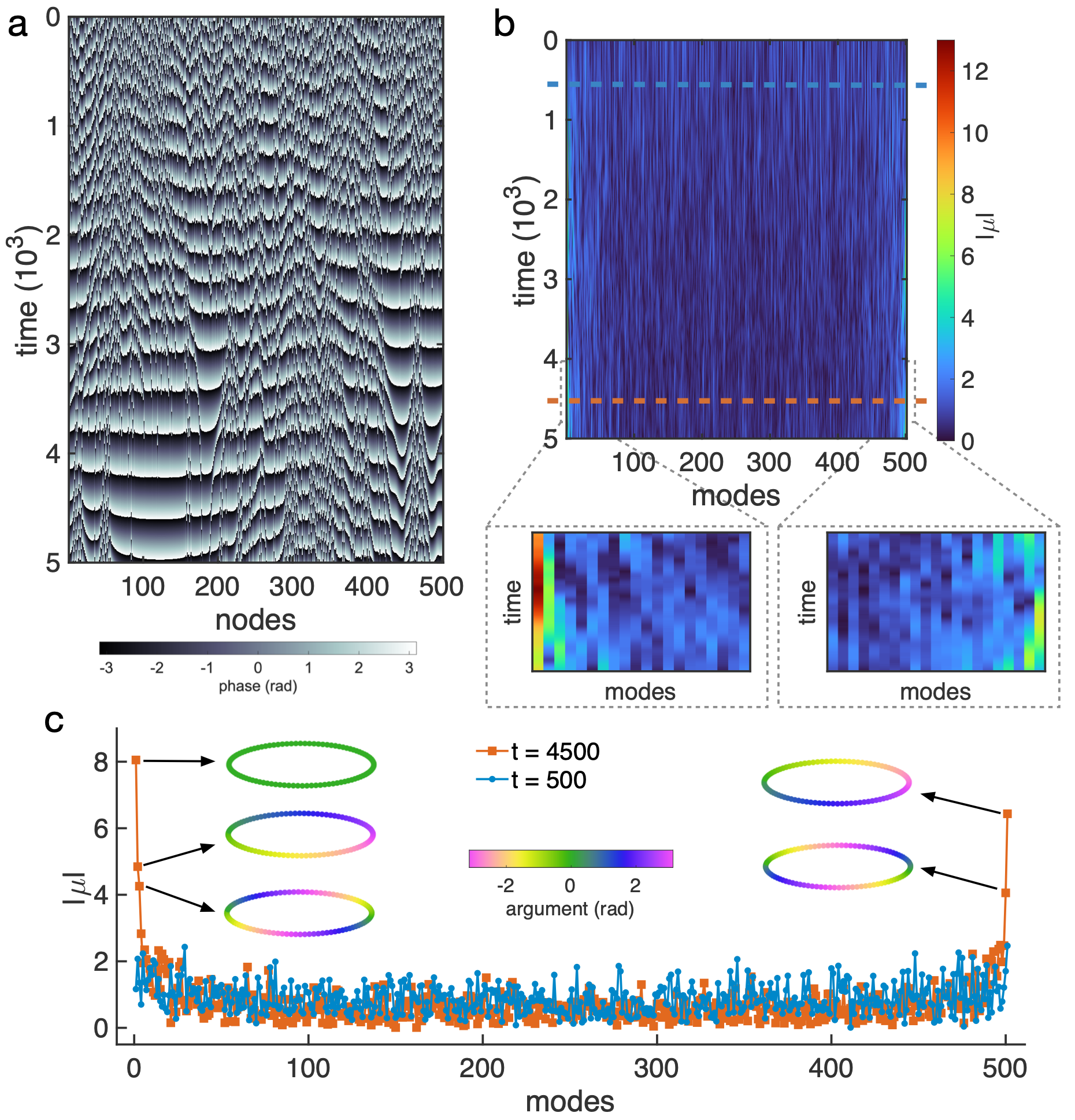}
    \caption{\textbf{A \revision{substrate} for \revision{transient} chimera states.} \textbf{(a)} We observe the emergence of chimera states in the network. \textbf{(b)} The eigenmode contribution, at first, appears to be uniform, however, a close look reveals a different pattern when the chimera state emerges (see inset). \textbf{(c)} When the network is asynchronous, at \revision{$t = 500$}, the modes contribution $|\mu|$ is uniform (blue line with dots); at \revision{$t = 4500$}, when a chimera state is observed, there are specific modes with higher contribution (red line with squares). These modes are related to eigenvectors with specific phase configuration (see inset): zero phase-difference and low spatial frequency waves traveling in opposite directions. Here, we consider $\epsilon = 20$ and $\phi = 1.40$.}
    \label{fig:chimera_modes}
\end{figure}

We then study \revision{the emergence of transient} chimera states using this approach. When each node in a network is connected by the same connectivity rule, such as with the complete graph, ring graphs, and distance-dependent networks, we can obtain the eigenvalues and eigenvectors of the system analytically via the circulant diagonalization theorem (CDT) \cite{davis1979} \revision{(see Supplemental Sec. II)}. \revision{The CDT states that the eigenvectors are given by $ (\bm{v}_{k})_s = (\sfrac{1}{\sqrt{N}}) \exp\left[ (\sfrac{-2\pi \i}{N}) (k-1)(s-1) \right]$, which are the columns of the discrete Fourier transform (DFT)}. \revision{Specifically,} the first eigenvector is $\bm{v}_{1} = [1, 1, \cdots, 1]$, and the arguments of its elements are $\mathrm{Arg}[\bm{v}_{1}] = [0, 0, \cdots, 0]$. In the context of our analysis, this eigenvector represents the zero phase-difference state. In a network arranged on a one-dimensional ring, the other eigenvectors have a phase configuration (or, argument of each element) representing waves with increasing spatial frequency. These eigenvectors appear in pairs that travel in opposite directions (clockwise and counter-clockwise; cf.~inset of Fig.\,\ref{fig:chimera_modes}c). Here, we quantify the contributions of each eigenvector in terms of the ``eigenmode contribution'' $\mu_{i}(t) = \langle \bm{x}(t), \bm{v}_{i} \rangle$, where $\langle . \rangle$ represents the complex inner product, and $i \in [1,N]$.

Figure \ref{fig:chimera_modes}a illustrates a chimera state in a network of \revision{Sakaguchi-}Kuramoto oscillators with distance-dependent \revision{power-law} connectivity \cite{note_matrix}. Due to random initial conditions, the network begins in an asynchronous state. The network then evolves to exhibit a chimera state, with clusters of synchronous and asynchronous nodes. Figure \ref{fig:chimera_modes}b shows $|\mu_{i}(t)|$ across time. When the network is an asynchronous state (\revision{$t = 500$}, blue dotted line, Fig.\,\ref{fig:chimera_modes}b), the eigenmode contribution is uniform across modes (blue line and dots, Fig.~\ref{fig:chimera_modes}c). When the network exhibits a chimera (\revision{$t = 4500$}, red dotted line, Fig.~\ref{fig:chimera_modes}b), however, the eigenmode contribution is no longer uniform:~the $\mu_{i}(t)$ are increased for the lowest and highest modes relative to the asynchronous state (red line and squares, Fig.\,\ref{fig:chimera_modes}c). This result indicates that, during \revision{transient} chimera states, the network exhibits a combination of low-spatial-frequency modes, traveling in both the clockwise and counter-clockwise directions around the ring, in addition to the mode representing zero phase offset $\mu_{1}$ (inset of Fig.~\ref{fig:chimera_modes}c). 
\begin{figure}[bht]
    \centering
    \includegraphics[width=\columnwidth]{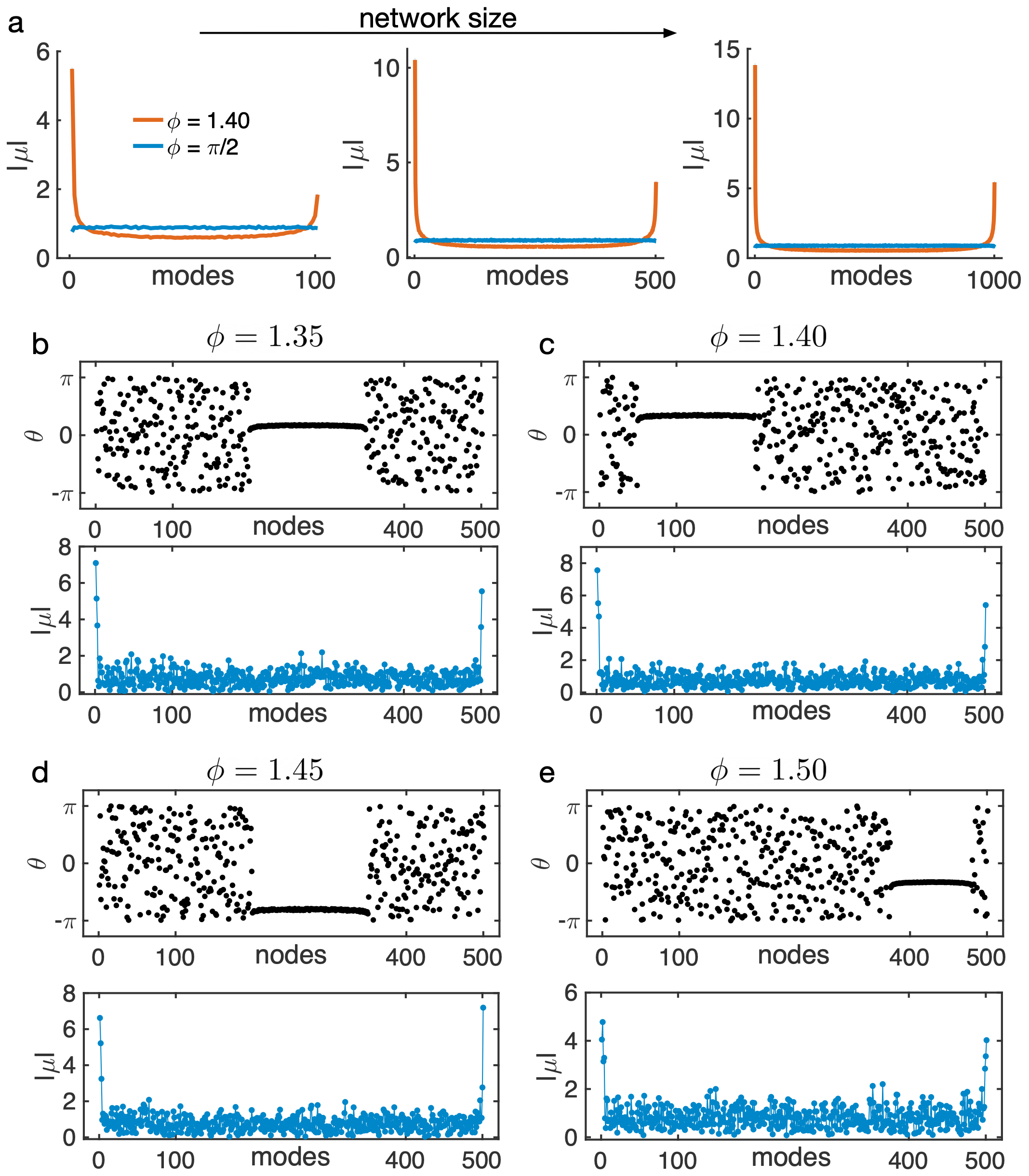}
    \caption{\textbf{Eigenmode contribution for finite networks.} \textbf{(a)} We analyze the eigenmode contribution $|\mu|$ for the final state of the network (\revision{$t = 5000$}). For $\phi = 1.40$ (red line) chimera states are possible, and the contribution of the first and last modes increase in the same manner as observed in Fig.\,\ref{fig:chimera_modes}c. As a control case, we consider $\phi = \sfrac{\pi}{2}$ (blue line) where the network remains in an asynchronous state and the eigenmode contribution is uniform across modes. These results consider $1000$ realizations with different initial conditions, where we plot the average with shaded area representing the standard mean error. We also consider different examples of chimera patterns for $N = 501$ and \textbf{(b)} $\phi = 1.35$, \textbf{(c)} $\phi = 1.40$, \textbf{(d)} $\phi = 1.45$, \textbf{(e)} and $\phi = 1.50$, where we observe that, in all cases, the modes contribution follows the similar fashion as shown in Fig.~\ref{fig:chimera_modes}c. Here, we consider $\epsilon = 20$.}
    \label{fig:modes_size_example}
\end{figure}

This analysis suggests a potential \revision{substrate} for \revision{transient} chimera states in \revision{finite Sakaguchi-}Kuramoto oscillator networks as a specific combination of modes. To verify that this is the case, we next studied a set of networks with increasing numbers of nodes across many different initial conditions. We consider three distance-dependent networks \cite{note_matrix} with size $N = 101$, $N = 501$, and $N = 1001$ and plot $|\mu|$ for the final state of the system at \revision{$t = 5000$}. Figure\,\ref{fig:modes_size_example}a illustrates the case with phase-lag ($\phi = 1.40$, red line), where the network displays chimera states, and with neutral coupling ($\phi = \sfrac{\pi}{2}$, blue line), where the network remains in an asynchronous state. When the network is asynchronous, the $|\mu|$ are uniform, demonstrating equal contribution across all eigenmodes. When the network exhibits a chimera state, however, the contribution of the first and last eigenmodes increases in the same manner as in Fig.\,\ref{fig:chimera_modes}. This result demonstrates that the \revision{condition} identified in Fig.\,\ref{fig:chimera_modes} occurs in the average over one thousand simulations with different, random initial conditions. Further simulations across different phase-lag values demonstrate that the identified pattern of eigenmode contributions holds across the range of parameters that lead to \revision{transient} chimera states in these networks, and for conditions where the synchronized cluster appears at different \revision{positions} in the network (Figs.\,\ref{fig:modes_size_example}b, \ref{fig:modes_size_example}c, \ref{fig:modes_size_example}d, and \ref{fig:modes_size_example}e). \revision{It is important to note that our approach generalizes to a diversity of networks, where the eigenspectrum of matrices that are not circulant can be obtained numerically (see Fig. S2 in the Supplement).} Taken together, these results provide further evidence that a specific combination of modes occurs generally, across initial conditions \revision{and networks}, when \revision{transient} chimera states appear.

\begin{figure}[tbh]
    \centering
    \includegraphics[width=\columnwidth]{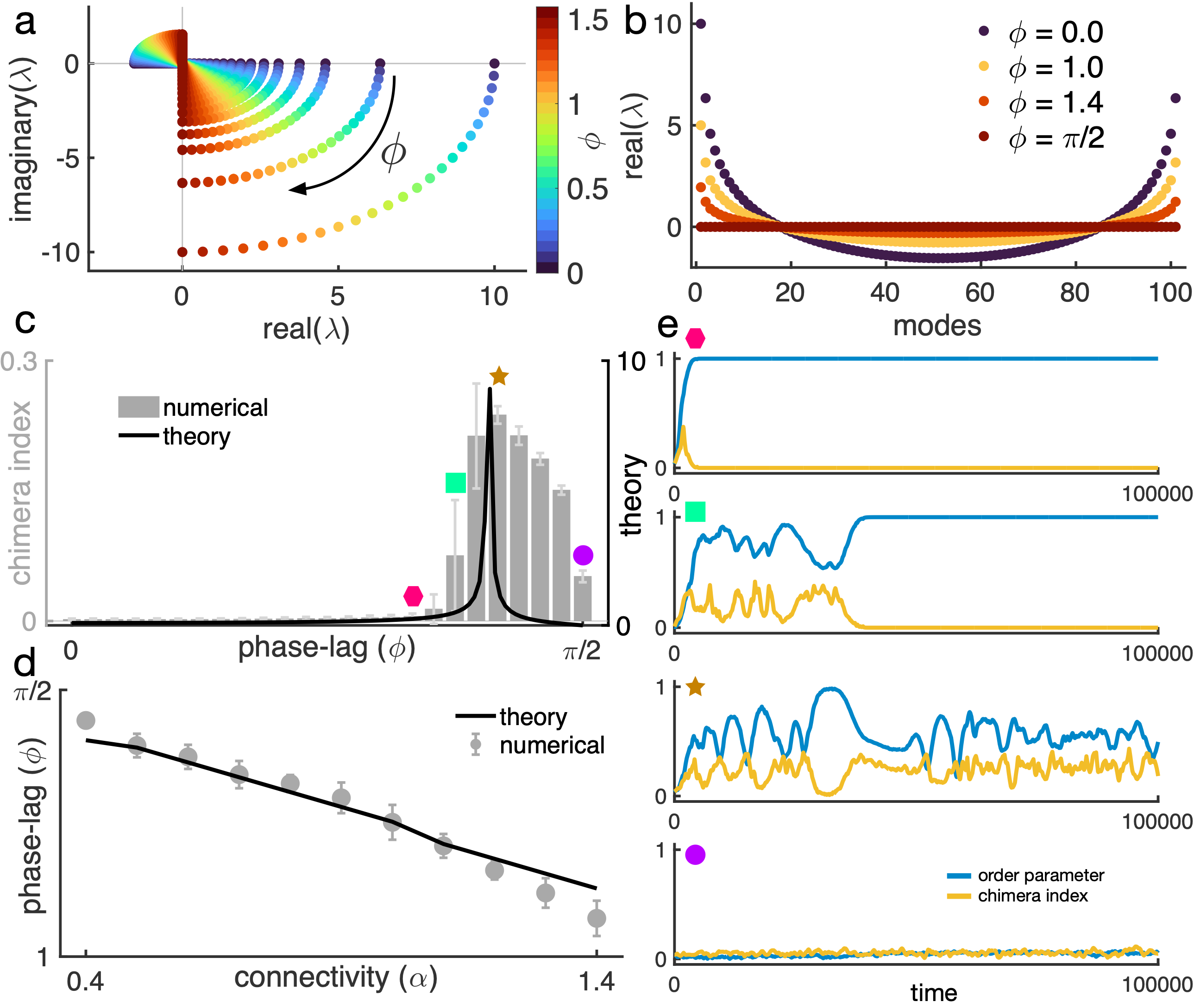}
    \caption{\textbf{Phase-lag and the emergence of \revision{transient} chimera states.} \revision{\textbf{(a)} As the phase-lag increases, the eigenvalues of $\bm{K}$ rotate with an angle $- \phi$ in the complex plane going from being purely real (for $\phi = 0$) to purely imaginary (for $\phi = \sfrac{\pi}{2}$).} \textbf{(b)} This rotation decreases the difference in the real part of the eigenvalues, which allows the network to access the modes relevant for the appearance of \revision{transient} chimera states. \revision{\textbf{(c)} We use a chimera index $\overline{\sigma}$ \cite{numerical_chi} for the numerical simulations (gray bars), where we observe that chimeras appear for a specific range of phase-lag $\phi$. These results are given by an average over $100$ simulations of $t = 10^{6}$ timesteps with errorbars indicating the standard deviation. We then use the spectral properties of $\bm{K}$ through Eq. (\ref{eq:theory_chi}) to analytically study this (black line). \textbf{(d)} This approach generalizes for networks with different connectivity (given by different values of $\alpha$), where our theory (black line) is able to capture the value of phase-lag where the chimera index is maximum (gray dots). \textbf{(e)}, Lastly, we show examples of the synchronization level (blue line) and the chimera index (yellow line) as a function of time for different phase-lag values based on panel (c) -- each represented by a different symbol.}}
    \label{fig:eigenvalues}
\end{figure}

What network mechanisms lead to this specific combination of modes, and thus to chimera states? To answer this question, we look to the eigenvalues of $\bm{K}$ to understand how these specific combinations of modes arise. Figure \ref{fig:eigenvalues}a shows the eigenvalues of $\bm{K} = \epsilon e^{-\i \phi} \bm{A}$ for increasing values of $\phi$. Our analytical approach allows us to understand the effect of $\phi$ as a rotation of these eigenvalues in the complex plane. When $\phi=0$, the eigenvalues of $\bm{K}$ fall along the real axis (blue dots, Fig.\,3a). When $\phi=\sfrac{\pi}{2}$, the eigenvalues fall along the imaginary axis (red dots, Fig.\,\ref{fig:eigenvalues}a). This rotation reduces the real part of the $\lambda_i$ that were initially positive, reduces the real part of $\lambda_1$ relative to the other eigenvalues, and - for the range of $\phi$ where chimera states are produced - the rotation leaves a small set of modes making a significant contribution to the dynamics (Fig.\,\ref{fig:eigenvalues}b). \revision{For all cases with $\phi < \sfrac{\pi}{2}$, however, the leading eigenvalue (in real part), is given by $\lambda_{1}$ which is associated with the phase synchronizing mode. Our approach thus analytically explains previous numerical observations that chimeras are transient states that are followed by phase synchronization in finite networks of Sakaguchi-Kuramoto oscillators \cite{wolfrum2011chimera}.}

\revision{Our theory offers further analytical insights into the role of phase-lag and connectivity on the emergence of transient chimeras. First, we use a chimera index $\overline{\sigma}$ \cite{numerical_chi} introduced in \cite{shanahan2010metastable} to measure the existence of chimera states in numerical simulations (see Section III in the Supplement). We consider a distance-dependent network with $\alpha = 1.0$ and vary the phase-lag $\phi \in [0, \sfrac{\pi}{2}]$. Consistent with previous observations \cite{shanahan2010metastable}, chimera states appear for a specific range of phase-lag (gray bars, Fig.~\ref{fig:eigenvalues}c). We then consider an analytical quantifier, based only on the spectral properties of $\bm{K}$, to study this phenomenon:}
\begin{equation}
    \revision{\zeta(\alpha,\phi) = {\Big|\lambda_{1}^{\ast} - \frac{1}{{\Delta_{1,2}}^{\ast}} \Big|}^{-1},}
    \label{eq:theory_chi}
\end{equation}
\revision{where \footnotesize{$\lambda_{1}^{\ast} = \eta(\alpha) \frac{\mathrm{Real}[\lambda_1(\alpha,\phi)]}{\mathrm{Real}[\lambda_{1}(\alpha,0)]}$} \normalsize{and} \footnotesize{${\Delta_{1,2}}^{\ast} = \frac{\mathrm{Real}[\lambda_{1}(\alpha,\phi) - \lambda_{2}(\alpha,\phi)]}{\mathrm{Real}[\lambda_{1}(\alpha,0) - \lambda_{2}(\alpha,0)]}$} \normalsize{(see Supplemental Sec. IV), with $\eta(\alpha)$ being the normalization factor for the distance-dependent network \cite{note_matrix}. This quantifier captures the balance between the first mode and the low spatial frequency modes, and thus expresses the emergence of chimera states. By using only two eigenvalues of $\bm{K}$, we are able to capture the non-monotonic behavior as a function of $\phi$ (black line, Fig.~\ref{fig:eigenvalues}c). This behavior has been numerically observed in \cite{shanahan2010metastable}, and our theory now offers an analytical explanation for it. This result generalizes for networks with different connectivity (different values of $\alpha$), where we can use our theory to analytically predict the precise phase-lag value where the chimera index is maximum (Fig.~\ref{fig:eigenvalues}d). Lastly, to illustrate the complexity of this phenomenon, we plot the order parameter and the chimera index as a function of time (blue lines and yellow lines, Fig.~\ref{fig:eigenvalues}e, respectively) for different values of $\phi$ based on Fig.~\ref{fig:eigenvalues}c (represented by different symbols). We observe that if the phase-lag is too small, the network quickly reaches phase synchronization (pink hexagon, Fig.~\ref{fig:eigenvalues}e). As the phase-lag increases, the transient becomes longer and chimera states appear (green square and brown star, Fig.~\ref{fig:eigenvalues}e). If the phase-lag is too close to $\sfrac{\pi}{2}$, however, the network remains in an asynchronous state and no chimera state is observed (purple circle, Fig.~\ref{fig:eigenvalues}e), and the order parameter and chimera index fluctuate due to the finite-size effects.}}

\begin{figure}[b!]
    \centering
    \includegraphics[width=\columnwidth]{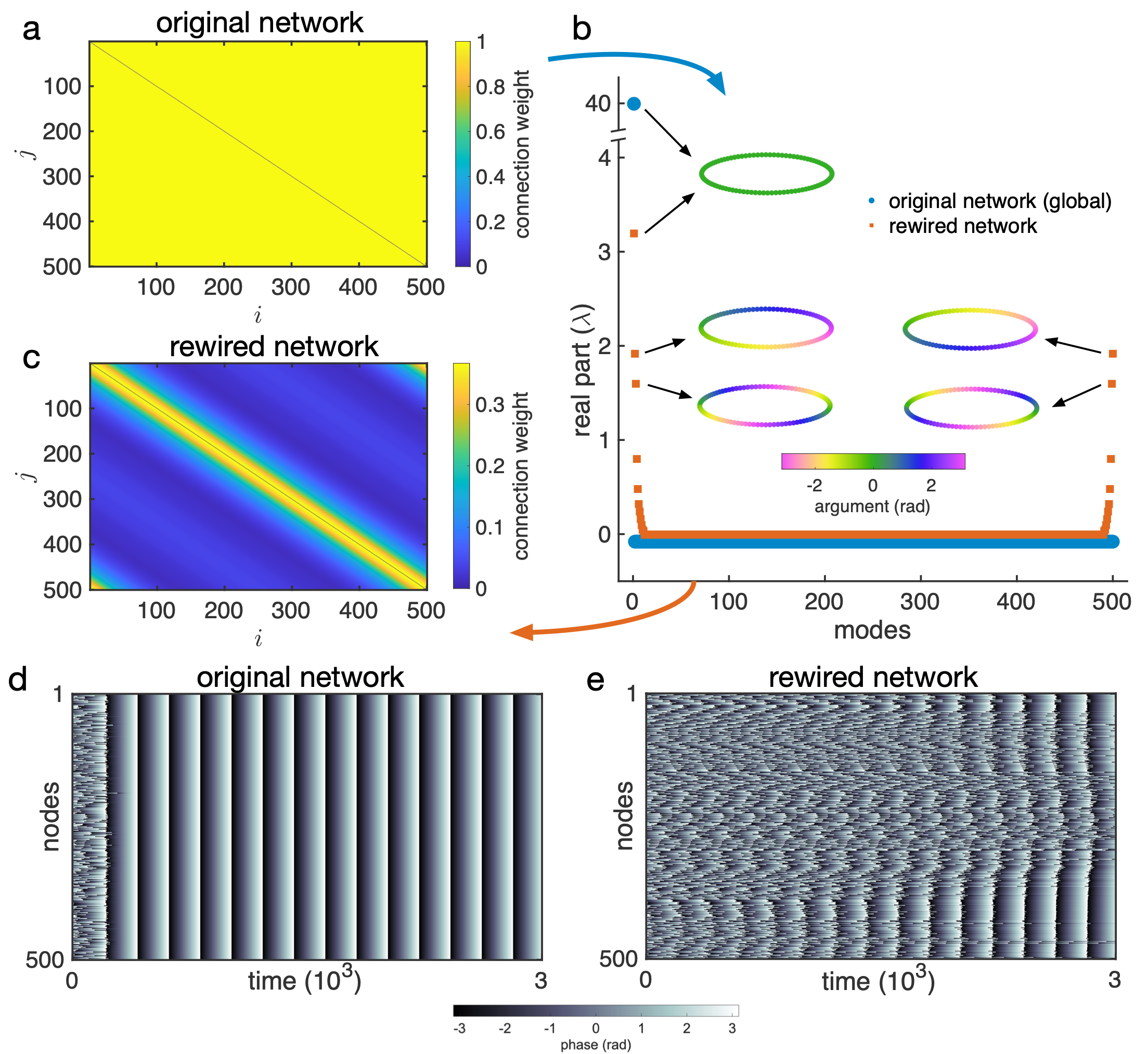}
    \caption{\textbf{Designing networks to exhibit chimeras.} \textbf{(a)} We start with a global network, which adjacency matrix $\bm{A}$ is given by the complete graph. \textbf{(b)} We use the CDT to obtain the eigenvalues of the matrix $\bm{K} = \epsilon e^{-\i\phi} \bm{A}$ for this network (blue dots). We then manually modify these eigenvalues in order to obtain a similar fashion as \revision{revealed by our approach in Fig. \ref{fig:eigenvalues}} with the modes relevant for chimera states having a higher real part in $\lambda$ (insets \textbf{b}). \textbf{(c)} We build the rewired network, where the adjacency matrix now has weighted connections. We then numerically integrate Eq. (\ref{eq:km_phase_lag}) considering: \textbf{(d)} the global network, where the system transitions to phase synchronization; \textbf{(e)} the rewired network with the same initial conditions, where the system exhibits chimera states. Here, $\epsilon = 0.08$ and $\phi = 0$.}
    \label{fig:designing_network_chimera}
\end{figure}
We now show that we can ``rewire'' a network to exhibit the eigenvalue configuration in Fig.\,\ref{fig:eigenvalues} and then express chimera states. To do this, we start with a globally connected network (the complete graph on $N$ nodes, $A_{ij} = 1$ for all $i \in [1, N], i \neq j$; Fig.\,\ref{fig:designing_network_chimera}a) and with no phase-lag ($\phi = 0$). We \revision{obtain} the eigenvalues and eigenvectors of $\bm{K} = \epsilon e^{-\i\phi} \bm{A}$ using the CDT, as above (blue dots, Fig.\,\ref{fig:designing_network_chimera}b). In the complete graph, only $\lambda_1$ has positive real part (Fig.\,\ref{fig:designing_network_chimera}b). Our analytical approach predicts this will lead the network to phase synchrony because $\lambda_1$ is associated with the zero phase-difference mode (inset, Fig.\,\ref{fig:designing_network_chimera}b). We then manually adjusted the eigenvalues to exhibit \revision{a configuration that our approach reveals to give rise to chimera states, as shown in Fig.~\ref{fig:eigenvalues},} and then obtained a ``rewired'' matrix $\bm{A}' = \bm{V} \bm{D}' \bm{V}^{-1}$ (Fig.\,\ref{fig:designing_network_chimera}c) with the new eigenvalues in $\bm{D}'$. In this case, while a numerical simulation of the original Kuramoto network connected by matrix $\bm{A}$ (complete graph) transitions to synchrony (Fig.\,\ref{fig:designing_network_chimera}d), the network with the new matrix $\bm{A}'$ now generates a chimera (Fig.\,\ref{fig:designing_network_chimera}e). This result demonstrates a specific analytical procedure for designing networks to produce chimera states in Kuramoto networks. \revision{This topic has received increasing attention in the past years with development of new algorithms to create specific networks to produce a desired pattern \cite{forrow2018functional, menara2022functional, masoliver2022embedded}, and} our analytical approach could have important impact for future machine learning algorithms for training neural networks to produce chimeras and other patterns.

\revision{There has been much recent interest in the possibility that neural systems exhibit chimera states, from the dynamics of spiking neural networks \cite{majhi2019chimera}, to the oscillations associated with memory storage and retrieval \cite{masoliver2023hippocampal}, and to the rhythms of sleep \cite{parastesh2021chimeras}. As a specific example, the 0.1-1 Hz slow oscillation that occurs in the deepest stages of non-rapid-eye-movement (non-REM) sleep has been observed to occur only in one hemisphere of the brain at a time, while the other half of the brain remains de-synchronized \cite{rattenborg1999half,mascetti2016unihemispheric,tamaki2016night,rattenborg2016evidence}. Explaining how one half of the brain can remain desynchronized while the other half exhibits highly synchronized neuronal activity has suggested connections to chimera states in oscillator networks \cite{masoliver2022embedded}. In uni-hemispheric sleep, synchronization occurs across networks of 50-100 cortical regions. While many mathematical approaches have provided insight into the origin of chimera states and provided bifurcation analyses \cite{abrams2004chimera,panaggio2015chimera}, our mathematical approach provides specific insight into the spatiotemporal impact of the eigenvectors of the network adjacency matrix in shaping these dynamics. This analytical insight links the eigenspectrum of the graph adjacency matrix to the resulting nonlinear dynamics in Sakaguchi-Kuramoto networks. This is, in general, a very difficult problem in applied mathematics \cite{curto2019relating}, but our analysis allows us to make this connection in this case. A key feature of our mathematical approach is to analyze the equations defining the network dynamics at the finite scale, which in turn allows us to understand the emergence of transient chimeras at the scale of tens to hundreds of nodes that is relevant for understanding neural dynamics during sleep.}


\revision{Our results demonstrate how this operator-based approach to networked systems at the finite scale can provide deep insight into the interplay of connectivity and nonlinearity at individual network nodes. Here, we have provided analytical insight into the interaction of local connectivity - including distance-dependent power-law graphs, ring graphs, clustered networks, and Watts-Strogatz graphs - and a phase-lagged sine interaction function to produce transient chimera states. Our theory provides a simple and precise mathematical formulation of how eigenvectors of the adjacency matrix combine to produce transient chimera states in finite Sakaguchi-Kuramoto dynamics (Figs.\,\ref{fig:chimera_modes} and \ref{fig:modes_size_example}), the interplay between phase-lag and network connectivity that leads to the emergence of these states (Fig.\,\ref{fig:eigenvalues}), and even how to design patterns of connectivity to produce chimera states (Fig.\,\ref{fig:designing_network_chimera}). Previous work has observed connections between the eigenvalues of the Laplacian matrix and phase synchronization in oscillator networks \cite{ocampo2021non,skardal2014optimal,mcgraw2007analysis}. Our theoretical approach provides a precise connection between network spectra and dynamics in two ways. First, our approach focuses on the network adjacency matrix, which is more directly related to the network dynamics, instead of the Laplacian. Second, our approach connects both the eigenvectors of the adjacency matrix and their corresponding eigenvalues to the resulting dynamics (specifically, when analyzing the matrix $\bm{K}$).}

\revision{Lastly, this work opens new avenues in the study of sophisticated spatiotemporal dynamics in oscillator networks. Future work can leverage our analytical predictions to control chimera states, a topic which has been studied with interest via numerical simulations until now \cite{bick2015controlling,sieber2014controlling}, or to study and control transient dynamics, which has also received much interest \cite{lilienkamp2020susceptibility,omel2022focusing}, because of its empirical relevance \cite{rabinovich2008transient,brinkman2022metastable,ocampo2021non}. Taking all these points together, it is possible to utilize chimera states and the analytical framework introduced here, to design controllable and interpretable dynamics that can advance our understanding of computation with spatiotemporal dynamics in nonlinear systems \cite{budzinski2023exact}.}

\begin{acknowledgments}
This work was supported by BrainsCAN at Western University through the Canada First Research Excellence Fund (CFREF), the NSF through a NeuroNex award (\#2015276), the Natural Sciences and Engineering Research Council of Canada (NSERC) grant R0370A01, Compute Ontario (computeontario.ca), Digital Research Alliance of Canada (alliancecan.ca), and the Western Academy for Advanced Research. R.C.B gratefully acknowledges the Western Institute for Neuroscience Clinical Research Postdoctoral Fellowship.
\end{acknowledgments}

%

\end{document}